\documentclass[floats,floatfix,showpacs,amssymb,prd,superscriptaddress,nofootinbib,
twocolumn,aps]{revtex4-1}
\usepackage{graphicx, epsfig, amssymb} %include figure files
\usepackage{amsmath, amsfonts}
\usepackage{bm} %include bold math: \bm{} creates bold letters in math mode

\usepackage[linktocpage]{hyperref}
\usepackage[caption=false]{subfig}
\usepackage[usenames]{color}

\def\be{\begin{equation}}
\def\ee{\end{equation}}
\def\beq{\begin{eqnarray}}
\def\eeq{\end{eqnarray}}

\newcommand{\bea}{\begin{eqnarray}}
\newcommand{\eea}{\end{eqnarray}}
\newcommand{\ben}{\begin{enumerate}}
\newcommand{\een}{\end{enumerate}}
\newcommand{\bi}{\begin{itemize}}
\newcommand{\ei}{\end{itemize}}

\def\be{\begin{equation}}
\def\ee{\end{equation}}

\begin{document}

\title{Stationary scalar configurations around extremal charged black holes}

 \author{Juan~Carlos~Degollado}\email{jcdaza@ua.pt}
   \affiliation{
   Departamento de F\'\i sica da Universidade de Aveiro and I3N, 
   Campus de Santiago, 3810-183 Aveiro, Portugal.
 }
 
\author{Carlos~A.~R.~Herdeiro}\email{herdeiro@ua.pt}
   \affiliation{
   Departamento de F\'\i sica da Universidade de Aveiro and I3N, 
   Campus de Santiago, 3810-183 Aveiro, Portugal.
}

\begin{abstract}
We consider the minimally coupled Klein-Gordon equation for a charged, massive scalar
field in the non-extremal Reissner-Nordstr\"om background. Performing a frequency domain
analysis, using a
continued fraction method, we compute the frequencies $\omega$ for quasi-bound states. We observe
that, as the extremal limit for both the background and the field is approached,
the real part of the quasi-bound states frequencies $\mathcal{R}(\omega)$ tends to the
mass of the field and the imaginary part $\mathcal{I}(\omega)$ tends to zero, for any
angular momentum quantum number $\ell$. The limiting frequencies in this double extremal limit are
shown to correspond to a distribution of extremal scalar particles, at
stationary positions, in no-force equilibrium configurations with the background. Thus,
generically, these stationary scalar configurations are regular at the event horizon.  If, on the
other hand, the distribution contains scalar 
particles at the horizon, the configuration becomes irregular therein, in agreement
with no hair theorems for the corresponding Einstein-Maxwell-scalar field system.

 \end{abstract}


\maketitle
%%%%%%%%%%%%%%%%%%%%%%%%%%%
\section{Introduction}
%%%%%%%%%%%%%%%%%%%%%%%%%%%

Scalar test fields in black hole (BH) geometries do not
admit, generically,  
stationary configurations with an asymptotic decay and with real frequencies, i.e bound states.
This follows from the physical
requirement that only ingoing waves can exist at the horizon, therefore preventing a real
equilibrium configuration between the field and the BH. Consequently, the
configurations allowed in BH backgrounds are \textit{quasi}-bound states, for
which the frequencies are complex, with the imaginary part revealing a time dependence for
the states, signaling either their absorption or, in the case of superradiant
instabilities, their amplification by the BH \cite{Press:1972zz}. 

The profile and some of the physical properties of quasi-bound states diverge at the horizon. 
This is intimately related to the inability that BH backgrounds have to
accommodate, in a regular fashion, the
scalar field, as an exact stationary solution, a property
established by no-hair theorems \cite{Bekenstein:1995un,Mayo:1996mv}. But even with this
caveat such quasi-bound states are informative. For instance in \cite{Burt:2011pv},
performing numerical simulations and starting with regular initial data for a
scalar field around a Schwarzschild BH, there were found damped oscillating 
solutions with frequency  and decay rate described by the real and imaginary parts of
quasi-bound state frequencies. These decay rates can be very small \cite{Barranco:2012qs}
and thus long lived scalar field configurations could exist around BHs,
even though eternal and regular configurations are, in general, precluded by no-hair theorems.

In this work we provide an example in which scalar field configurations around a BH can become
stationary and an interpretation to justify why this is possible. We start by  finding the
quasi-bound states for a charged massive scalar field in a
non-extremal  Reissner-Nordstr\"om (RN) BH, and show that
these states have a simple limiting behaviour as extremality for both the BH and the field is
approached:  the imaginary
part of the frequency vanishes and the real part of the frequency tends to the test field mass (and
charge).
Then, the scalar field configurations obtained in this limit are understood as the electrostatic
potential of a
distribution of extremal scalar particles in equilibrium with the extremal
BH, thus providing examples of scalar fields around extremal charged BHs which are regular on the
event horizon. 
Such configurations do not preserve
spherical symmetry around the BH and that is the way they circumvents no-hair
theorems.

This paper is organized as follows. In Sec. \ref{wave_equation} we discuss solutions of the
minimally coupled Klein-Gordon equation on the RN background. Quasi-bound state solutions of this
equation are considered in more detail in Sec. \ref{sec:frequencies}, where some explicit
frequencies are computed and analyzed. In Sec. \ref{sec:extremal} the extremal limit is discussed,
by computing the states with frequencies equal to the limiting behaviour observed in Sec. 
\ref{sec:frequencies}, and an interpretation for these states is given. We draw some concluding
remarks and comment on the non-linear solution including the backreaction of the scalar field in
Sec. \ref{sec:conclusions}.

%%%%%%%%%%%%%%%%%%%%%%%%%%%%%%%%%%%%%%%%%%%%%%%%%%%%%%%%%%%%%%%%%%%%%%%%%%%%%%
\section{Background and test field}
\label{wave_equation}
%%%%%%%%%%%%%%%%%%%%%%%%%%%%%%%%%%%%%%%%%%%%%%%%%%%%%%%%%%%%%%%%%%%%%%%%%%%%%%
We consider a massive, charged scalar field, $\Phi$, with mass $\mu$ and charge
$q$, obeying the wave equation
\be
\left[\hat{D}^\nu \hat{D}_\nu-\mu^2\right]\Phi=0 \ ,
\label{we}
\ee
where $\hat{D}_\nu \equiv D_\nu-iqA_\nu$. This field is  propagating in the background of a
Reissner-Nordstr\"om BH with charge $Q$ and 
mass $M$:
\be
ds^2=-f(r)dt^2+\frac{dr^2}{f(r)}+r^2(d\theta^2+\sin^2\theta d\phi^2) \ , 
\ee
where $f(r)={(r-r_+)(r-r_-)}/{r^2} $, $r_\pm\equiv M\pm\sqrt{M^2-Q^2}$ 
and $A_\nu dx^\nu=-Q/r \, dt$. 
Taking the standard ansatz for the scalar field which reflects the spherical symmetry and 
staticity of the background:
\be
\Phi= \sum_{\ell,m}\Phi_{\ell}^m\equiv \sum_{\ell,m} e^{-i\omega
t}Y_\ell^m(\theta,\phi)R_{\ell}(r) \ ,
\ee
where $Y^m_\ell$ are the spherical harmonics and $\omega$ the \textit{complex} 
frequency of a wave, (\ref{we}) yields the radial equation for each mode:
\be
r^2 f\frac{d}{dr}\left(r^2 f \frac{dR_\ell(r)}{dr}\right)+UR_\ell(r)=0 \ ,
\label{re0}
\ee
where $U= r^2\left[(\omega r-qQ)^2-f(\mu^2r^2+\ell(\ell+1))\right]$. 
Observe that the azimuthal quantum number $m$ is irrelevant due to spherical symmetry. 
The solution for the field $\Phi$ is immediately obtained by solving the radial equation
for each mode (\ref{re0}), due to the linearity of the wave equation (\ref{we}). In terms of $Z(r) =
r\,R(r)$, and dropping the subscript $\ell$ for notation simplicity, (\ref{re0}) becomes
\begin{eqnarray}
 \displaystyle{ \frac{d^2}{dr^2}Z(r)+\frac{f'}{f}
\frac{d}{dr}Z(r)}+\frac{1}{f^2}\left[\omega^2-V_{eff}(r)\right]Z(r)=0 \ , \ \ \ \  \label{eq:sl} 
\end{eqnarray}
where we have defined the  effective potential
\begin{equation}
V_{eff}(r)= \frac{2qQ\omega}{r}-\frac{q^2Q^2}{r^2}+f\left(
\frac{l(l+1)}{r^2}+\mu^2+\frac{f'}{r} \right) \ .
\end{equation}
Alternatively, introducing the Regge-Wheeler radial coordinate $r^*$, by $dr^*=dr/f(r)$,
this wave equation is rewritten as 
\be
\left[-\frac{d^2}{dr^{*2}}+V_{eff}(r)\right]Z(r)=\omega^2Z(r) \ ,
\label{werstar}
\ee
where $r=r(r^*)$. The properties of the potential $V_{eff}(r)$ have been discussed in the past, see
eg.~\cite{Furuhashi:2004jk}. In particular one can show that the height of the
centrifugal barrier increases with the charge of the field and that the constant value of
the
potential near the outer horizon also increases with the charge of the field but only up
to some maximum; then it starts decreasing. The main feature of this potential, however, is
that for a given combination of the parameters it exhibits a well, that can be considered
as one of the key ingredients to have quasi-bound states.

In order to solve the differential equation \eqref{eq:sl} we must provide a
set of suitable boundary conditions at the horizon and at spacial infinity.
To see the most relevant feature of the near horizon behaviour we note that in this
region equation (\ref{werstar}) becomes to leading order:
\be
\frac{d^2}{dr^{*2}}Z(r)+(\omega-q\phi_+)^2Z(r)\simeq 0  \ ,
\ee
where $\phi_+=Q/r_+$ is the electrostatic potential of the external  horizon. This 
equation is solved by a superposition of in and outgoing waves. Choosing the solution 
\be
Z(r)\stackrel{r\rightarrow r_+}{\sim} e^{-i(\omega-\omega_c)r^*} \ ,
\label{nh}
\ee
where $\omega_c\equiv q\phi_+$, corresponding to an ingoing wave for $q=0$, one observes 
the salient feature that it becomes an outgoing wave for $\omega< \omega_c$ (in this electromagnetic
gauge). This is the
condition for \textit{superradiance}.

Asymptotically, keeping the terms of order $1/r$ in equation \eqref{re0} one gets
\begin{equation}
 R(r)\stackrel {r\rightarrow \infty}{ \sim} \frac{e^{\chi r}}{r^{1-\sigma}} \ , 
 \label{rinf}
\end{equation}
where
\be
\sigma\equiv \frac{qQ\omega+M\mu^2-2M\omega^2}{\chi} \ , \qquad \chi\equiv  \pm
\sqrt{\mu^2-\omega^2} \ . 
\ee
From \eqref{rinf} one observes a qualitatively distinct behaviour depending on the sign of the
real part 
of $\chi$, $\mathcal{R}(\chi)$. In particular, for $\mathcal{R}(\chi)<0$ we have 
\textit{quasi-bound states}. These are characterized by a decaying behaviour at spatial infinity.
For
$\mathcal{R}(\chi)>0$ we have \textit{scattering states}.
Hereafter we will be interested in quasi-bound states. 

%%%%%%%%%%%%%%%%%%%%%%%%%%%%%%%%%%%%%%%%%%%%%%%%%%%%%%%%%%%%%%%%%%%%%%%%%%%%%%
\section{Semi-analytic global solution: Quasi-bound states frequencies}
\label{sec:frequencies}
%%%%%%%%%%%%%%%%%%%%%%%%%%%%%%%%%%%%%%%%%%%%%%%%%%%%%%%%%%%%%%%%%%%%%%%%%%%%%%
To find the solution of equation (\ref{eq:sl}) in the region $r>r_{+}$
we will use a continued-fraction procedure developed by Leaver to find
the quasinormal modes for the Schwarzschild and Kerr BHs
\cite{Leaver:1985ax}.\footnote{In \cite{Leaver:1990zz}, 
Leaver considers a similar method for a Regge-Wheeler equation obtained from perturbing
the Reissner-Nordstr\"om background.} This amounts to take a power series ansatz with a
pre-factor adapted to the boundary conditions observed in the previous section
\begin{equation}
 Z(r)=e^{\chi r}u^{\rho}(r-r_{-})^{\sigma-1}\,r\,\sum_{n=0}^\infty a_{n}u^{n}\ ,\label{eq:ansatz}
\end{equation}
where 
\begin{eqnarray}
& \displaystyle{u=\frac{r-r_{+}}{r - r_{-}}\,, \quad \quad \rho = -i\, r^{2}_{+}\left( 
\frac{\omega-\omega_c}{r_{+}-r_{-}} \right)    } \, . &
\end{eqnarray}
Substituting \eqref{eq:ansatz} into \eqref{eq:sl} we obtain a three term recurrence
relationship for the $a_n$ of the form:
\begin{eqnarray}
\label{eq:tterm}
& \alpha_{0}a_{1}+\beta_{0}a_{0} =0 \ ,& \\ 
& \alpha_{n}a_{n+1}+\beta_{n}a_{n}+\gamma_{n}a_{n-1}
=0\ ,  \quad n=1,2,3,...\ , & \nonumber
\end{eqnarray}
with
\begin{eqnarray*}
 \alpha_{n} &=& (1 - Q^2)n^2 + c_0n - (1 - Q^2) + c_0\  ,\\
 \beta_{n} &=&  -2(1 - Q^2)n^2   + c_1n + c_2 \  , \\
 \gamma_{n} &=&  (1 - Q^2)n^2 + c_3n + c_4\  .
\end{eqnarray*}
The constants $c_{i}$ are lengthy expressions but otherwise straightforward to obtain.
For a non charged massive scalar field on a Schwarzschild background, these
expressions reduce to the ones (in the same limit) given in Ref.
\cite{Dolan:2007mj}.
The procedure to find the frequencies of the quasi-bound states consists in setting the three
term
relationship \eqref{eq:tterm} in the form of a continued fraction algebraic equation and
then
solving it with a root finding procedure.

Furthermore, we have checked the value
of the frequencies obtained with the continued fraction method by numerically integrating
the radial equation. We took as initial condition the behaviour of the function close to
the horizon and integrated up to some large $r$ compared with the horizon radius.

\begin{figure}[ht]
\vspace{0.2cm}
\includegraphics[width=0.48\textwidth]{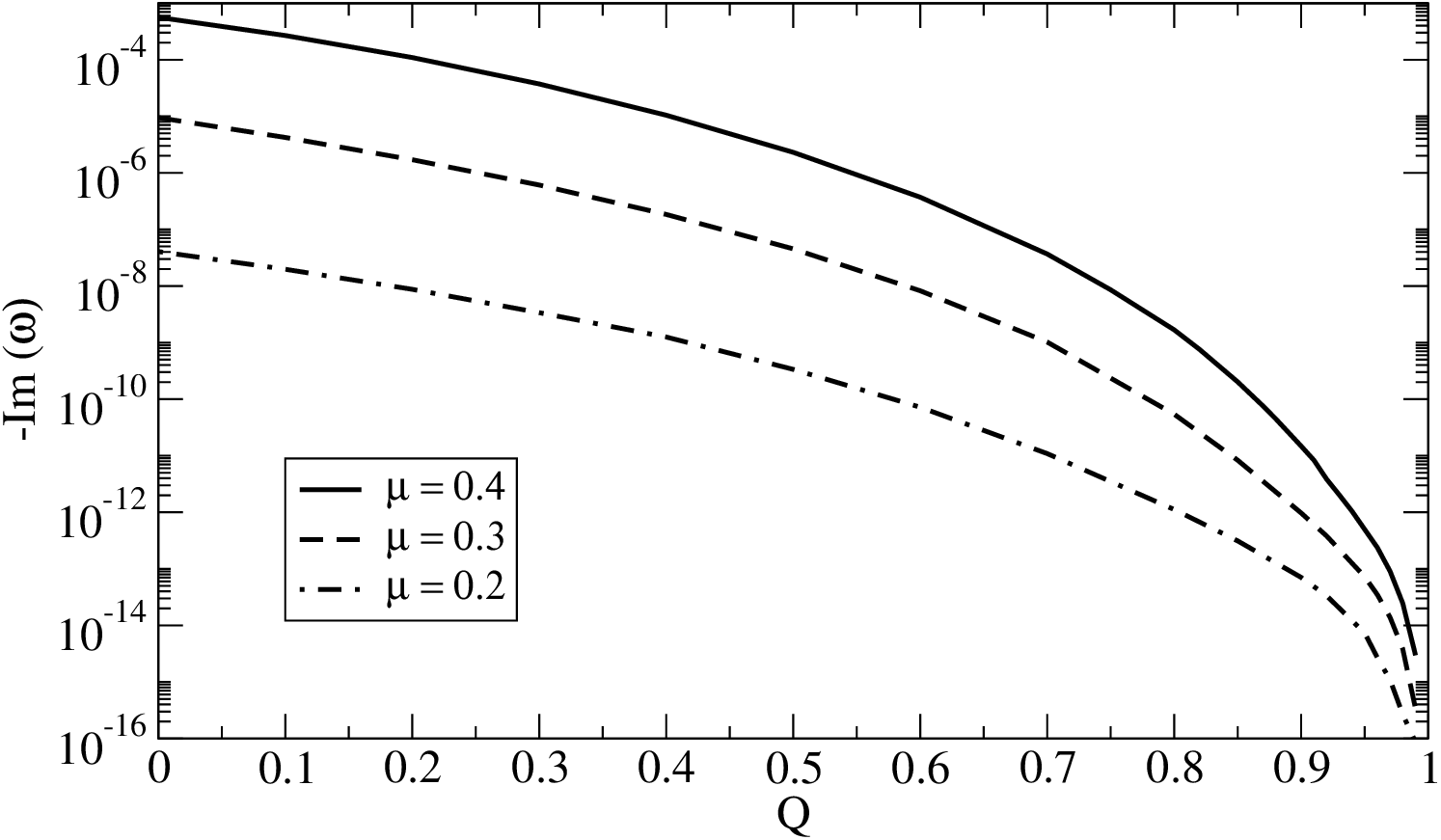} \,  
\includegraphics[width=0.48\textwidth]{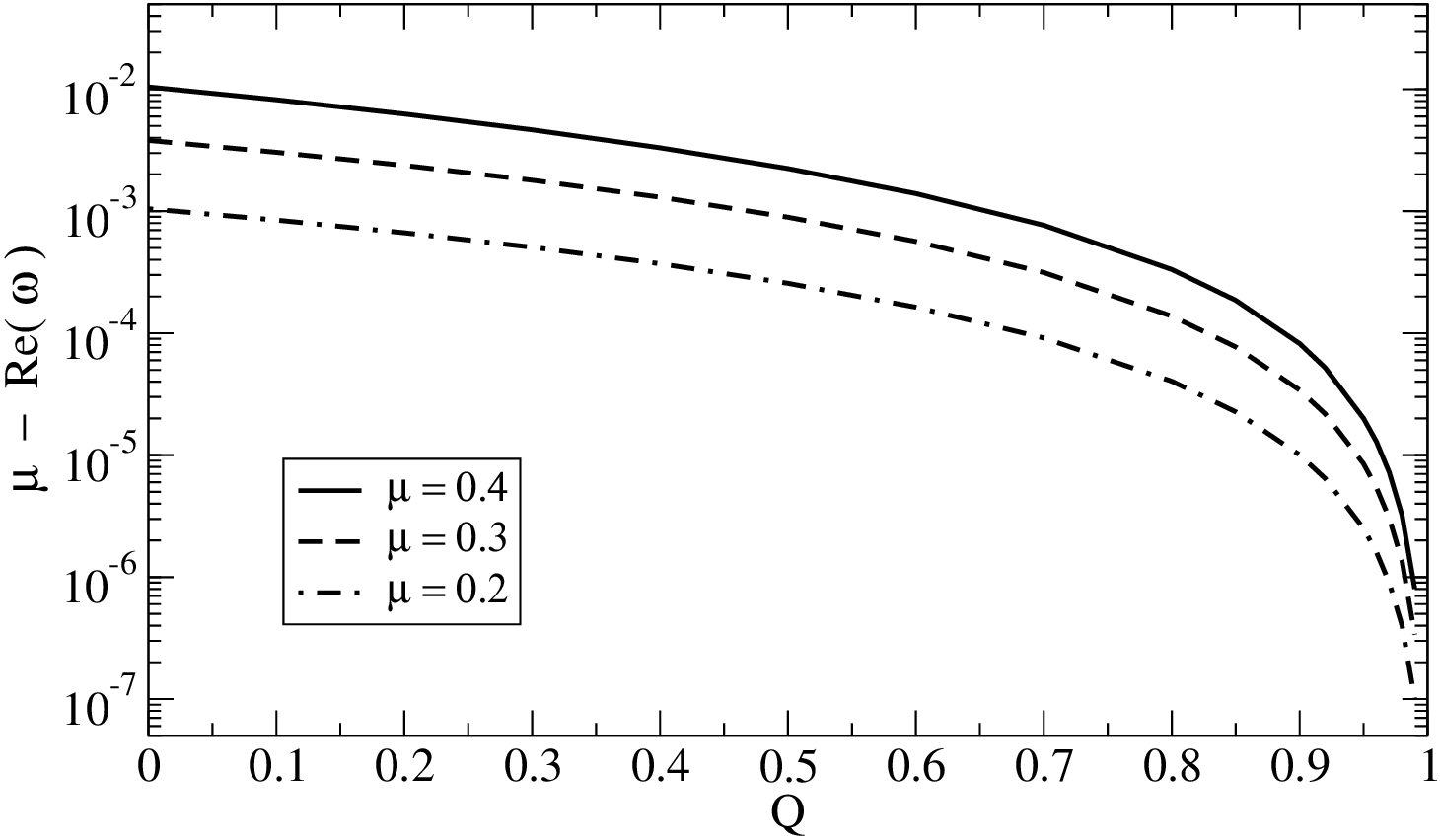}
\caption{(Top panel) The imaginary part of the frequency always tends to zero, even if the rates of
decay do depend on the value of $\mu$.  (Bottom panel) The real part of
the frequency tends to $\mu$ as $|Q|/M\rightarrow 1$. In both plots we keep $\mu/|q| = 1$, $\ell=1$
and $M=1$.}
\label{fig:Rew_vs_Q}
\end{figure}

The observed trend for the frequency of the quasi-bounded states is that, as the test field
becomes extremal
($\mu=|q|$), the imaginary part of the frequency decreases.
The imaginary part of the frequency is 
a measure of the rate at which the field falls into the
horizon. For the charged massive scalar field, the closer the background black
hole is to extremality the smaller this imaginary part is for the fundamental tone. The real part of
the frequency,
on the other hand, tends to increase and converge towards the field mass as the BH charge is
increased to extremality. This trend is shared by modes with different angular
momentum quantum number. To make clear the behaviour of both the real and imaginary parts of the
quasi-bound states
frequencies as the double extremality is attained ($|Q|,|q|\rightarrow M,\mu$), we plot
them against the BH charge for $q/\mu=1$ in Fig.~\ref{fig:Rew_vs_Q}. 
\begin{figure}[h]
\includegraphics[width=0.48\textwidth]{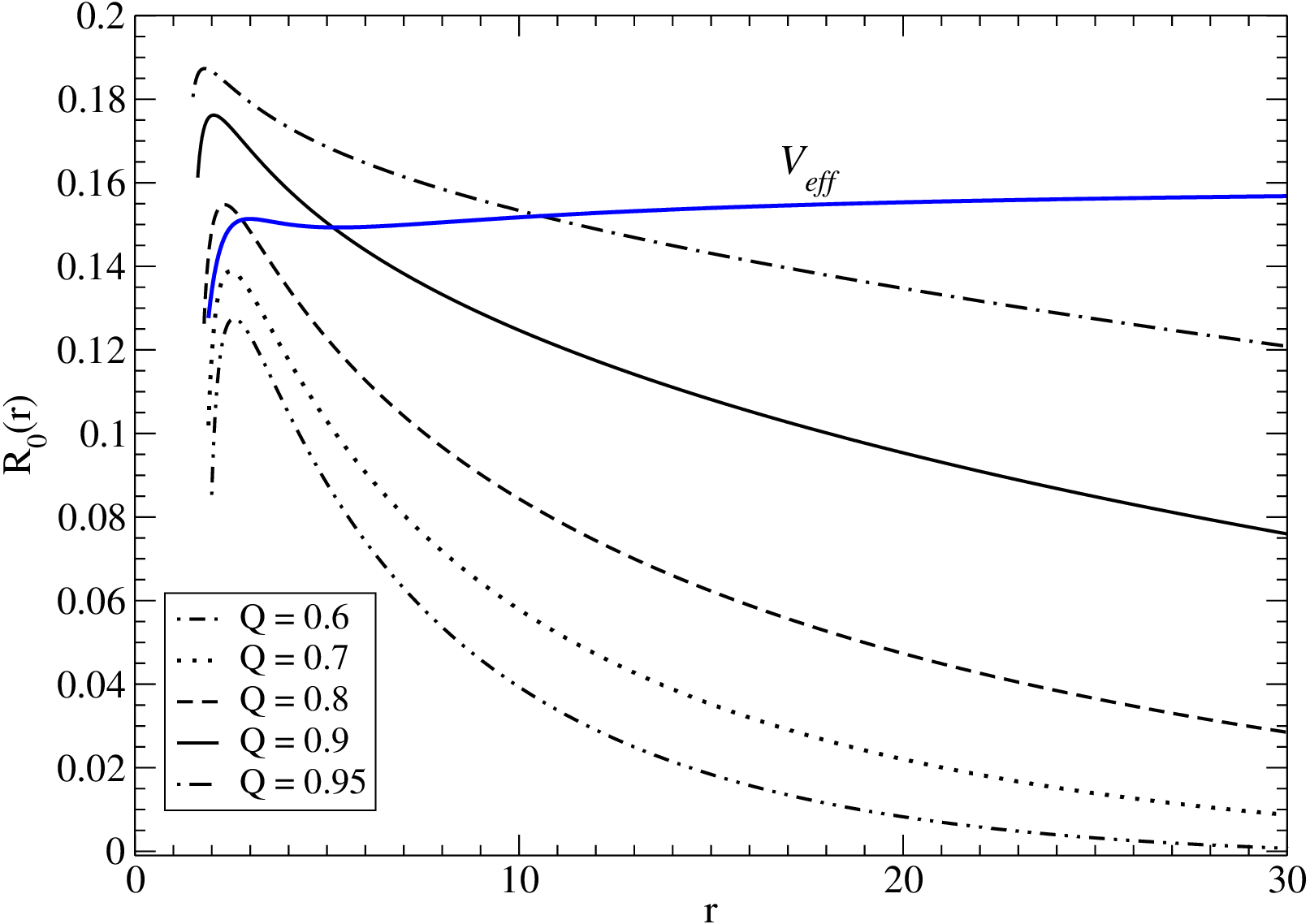}\vspace{0.45cm}
\includegraphics[width=0.48\textwidth]{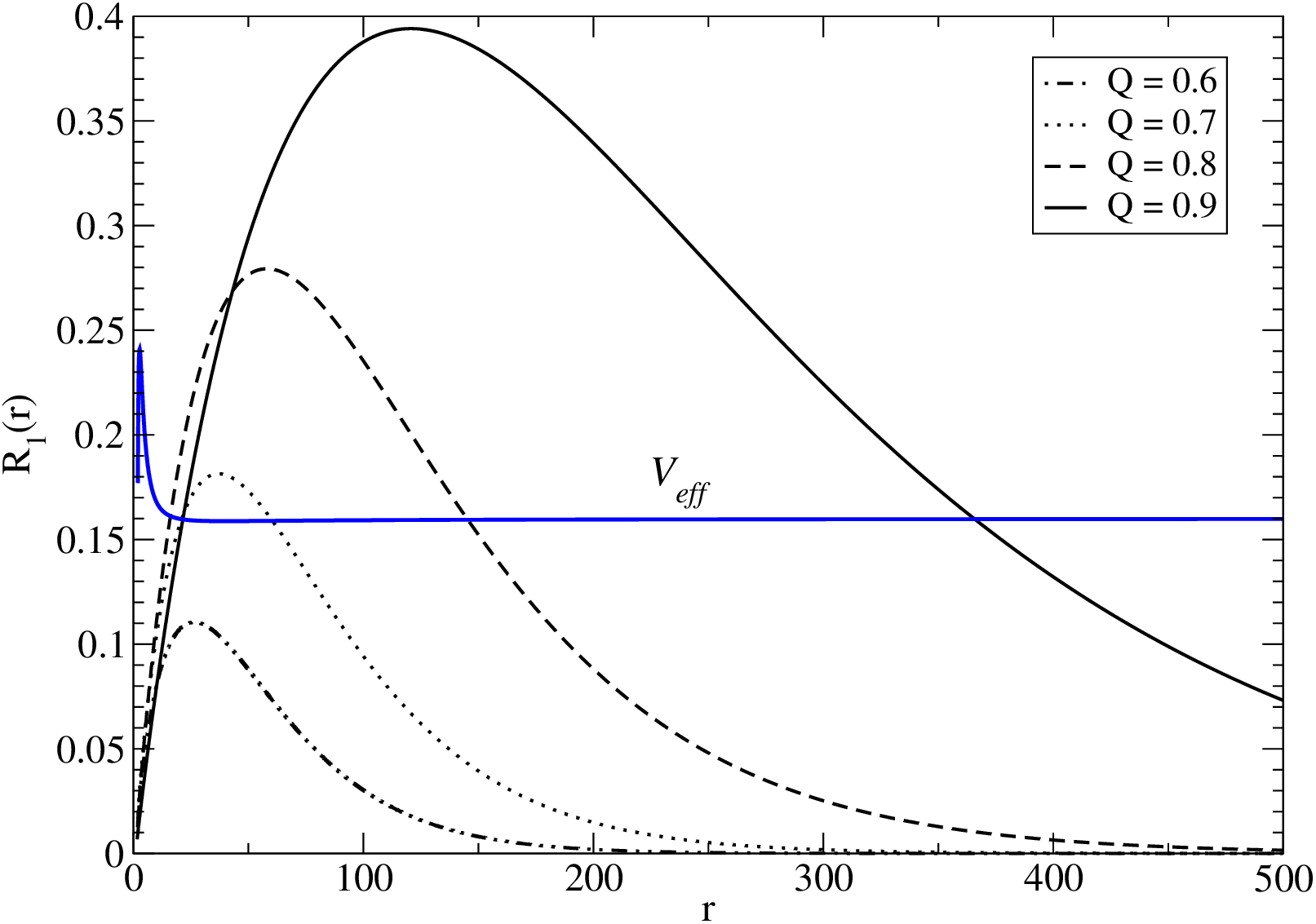}
\caption{The real part of the radial function for some values of the BH charge. We have also plotted
the effective potential for $Q=0.7$ and $q=\mu=0.4$. (Top panel) For $\ell = 0$ the maximum of the
radial function is approaching the outer horizon, as extremality of the background is approached.
(Bottom panel) For $\ell=1$ the function tends to spread along the potential well.}
\label{fig:profiles}
\end{figure}
Concerning the radial function as the double extremal limit is taken, the profile
of the $R_{0}(r)$
mode is qualitatively different from the profiles for the others modes. When $\ell=0$ ($s$-wave),
the radial
function is localized in the region between the external horizon and the
maximum of the potential barrier. As $|Q|/M\rightarrow 1$, $R_{0}(r)$ becomes
narrower in such a way that its maximum tends to the horizon. 
For $\ell\neq0$, the functions $R_{\ell}(r)$
tend to spread out as the double extremal limit is approached - Fig \ref{fig:profiles}. 

%%%%%%%%%%%%%%%%%%%%%%%%%%%%%%%%%%%%%%%%%%%%%%%%%%%%%
\section{Extremal black hole}
\label{sec:extremal}
%%%%%%%%%%%%%%%%%%%%%%%%%%%%%%%%%%%%%%%%%%%%%%%%%%%%%
The results of the previous section describe the limiting behaviour of the quasi-bound
states as the double extremal limit is attained. We shall now consider exactly this limit
by focusing on the extremal Reissner-Nordstr\"om BH,  $|Q|=M$ ($r_{\pm}=M$), with
an extremal test field
($\mu=|q|$). Then, using a new radial coordinate $\rho\equiv r-M$, for states with $\omega=\mu=|q|$
and $qQ>0$,
the radial wave equation \eqref{re0} reduces to 
\begin{equation}
\frac{d^2
R_{\ell}}{d\rho^2}+\frac{2}{\rho}\frac{dR_{\ell}}{d\rho}-\frac{\ell(\ell+1)}{\rho^2}R_{\ell}=0\ .
\label{eq:flat_lap}
\end{equation}
This equation is the radial part of the Laplace equation on Euclidean 3-space
$\mathbb{E}^3$
whose solution is $R_{\ell}=A_{\ell}\rho^{\ell}+{B_{\ell}}/{\rho^{\ell+1}}$.
The spacial part of the
scalar field $\Phi$ is therefore a linear combination of harmonic functions 
\begin{equation}
\Phi=e^{-i\mu t}\sum_{\ell,m}Y_\ell^m(\theta,\phi)
\left[A_{\ell}\rho^{\ell}+\frac{B_{\ell}}{\rho^{\ell+1}}\right] \ .
\label{harradial}
\end{equation}
Each of these partial waves, with appropriate $A_\ell,B_\ell$, describes the double extremal limit
of a quasi-bound state.

To interpret the meaning of the modes (\ref{harradial}) and understand their appearance,
it is useful to rewrite the extremal RN background using the coordinate $\rho$; this
corresponds to \textit{isotropic coordinates}. Then the fields take the form
\begin{equation}
 ds^2 = -H^{-2} dt^2 + H^2\delta_{ij}dx^idx^j\ , \qquad A=H^{-1}dt \ ,
 \label{mpform}
\end{equation}
where furthermore $\rho=\sqrt{\delta_{ij}x^ix^j}\equiv |{\bf x}|$ and $H$ is a harmonic
function on
Euclidean 3-space with a simple pole localised at the origin: $H=1+{M}/{|{\bf x}|}$. In these
coordinates ${\bf x}=0$ is the location of the extremal RN BH horizon.

Taking the scalar field in the form $\Phi(t,x^i)=e^{-i\mu t}\tilde{H}(x^i)$, the wave
equation (\ref{we})  with $\mu=|q|$ in the background (\ref{mpform}), yields the harmonic
equation $\Delta_{\mathbb{E}^3} \tilde{H}=0$. One solution is the harmonic function with a single
pole at ${\bf x}'$: $\tilde{H}={\mu}/{|{\bf x}-{\bf x}'|}$. This describes the electric potential of
one particle located at this pole. Expressed in terms of spherical coordinates on ${\mathbb{E}^3}$,
$(\rho,\theta,\phi)$, chosen such that ${\bf x}'$ lies  at coordinates  $(\rho',\theta'\neq
0,\phi')$, then
\begin{eqnarray}
\tilde{H}=\frac{\mu}{|{\bf x}-{\bf
x}'|}=\sum_{\ell,m}Y_\ell^m(\theta,\phi)\frac{B_\ell(\rho',\theta',\phi')}{\rho^{\ell+1}} \ ,
\label{equality}
\end{eqnarray}
where
\be
B_\ell(\rho',\theta',\phi')=\frac{4\pi \mu}{2\ell+1}  (\rho')^\ell Y_{\ell}^{*m}(\theta,'\phi') \ .
\label{19}
\ee
So $\tilde{H}$ is indeed the spatial part of \eqref{harradial}, with $A_\ell=0$. This fact shows
that the radial profiles (\ref{harradial}) correspond to partial waves
for an extremal point-like scalar source of mass and charge $\mu$ displaced from the BH horizon,
i.e. at $\rho\neq 0$. If, one the other hand, the particle is at $\rho=0$, it follows from
\eqref{equality}-\eqref{19} that only the $s$-wave appears.

There is an 
agreement between the behaviour of the $s$-wave seen in Sec. \ref{sec:frequencies} as extremality is
approached, and the one exhibited for the extremal case in this section. For the latter, a pure
$s$-wave corresponds to a source localised at
the horizon, whereas in the former, the peak of the radial function tends to the origin, as
displayed in the top panel of Fig. \ref{fig:profiles}. A related observation is that
the harmonic function $\tilde{H}$ is regular at the horizon except if the source is localised there.

A similar interpretation for the modes in \eqref{harradial} carries through if $\tilde{H}$
represents multiple scalar sources instead of a single one. 
We then have a superposition of harmonic functions with localised poles at fixed positions ${\bf
x}_k'$, corresponding to spherical coordinates $(\rho_k',\theta_k',\phi_k')$,
\be
\tilde{H}=\sum_k\tilde{H}_{k}=\mu \sum_k\frac{1}{|{\bf x}-{\bf x}'_{k}|} \ ,
\label{solh}
\ee
which may again be rewritten, in spherical coordinates, as the multipolar expansion in the right
hand side of \eqref{equality}, with $B_\ell(\rho',\theta',\phi')$ replaced by $\sum_k
B_\ell(\rho_k',\theta_k',\phi_k')$, corresponding to replacing one particle by many particles. 

The existence of stationary scalar states and their interpretation can, furthermore, be generalized
to a background with multiple extremal BHs instead of a single one. This is achieved replacing $H$
by a superposition of harmonic functions with localised
poles at different points, ${\bf x}_i$, $H=1+\sum_i {M_i}/{|{\bf x}-{\bf x}_i|}$, since the scalar
field equation still reduces to a harmonic equation on $\mathbb{E}^3$.
Such solution of the Einstein-Maxwell system is the well known Majumdar-Papapetrou multi
BH solution~\cite{Papapetrou1945,Majumdar:1947eu}, corresponding to a collection
of BHs with mass and charge $M_{i}=Q_{i}$, placed at arbitrary positions ${\bf x}_i$
\cite{Hartle:1972ya}, held in equilibrium by a balance between gravitational and
electrostatic forces. A multiple scalar particle configuration will be regular on each horizon of
the Majumdar-Papapetrou background as long as ${\bf x}_i\neq {\bf x}_k'$, for all $i,k$.

The scalar particles are in equilibrium with the BHs due to a `no-force'
condition, a balance between gravitational and electromagnetic forces, as can be easily
checked by studying the orbits generated by the Lagrangian
$\mathcal{L}=\mu\sqrt{-g_{\alpha\nu}\dot{x}^\alpha \dot{x}^\nu}+\mu A_\alpha
\dot{x}^\alpha$ in the background (\ref{mpform}), where `dot' denotes derivative with
respect to proper 
time. This Lagrangian is adequate to describe the interaction of the scalar particles
with the background \eqref{mpform} because, at linear level, there is no interaction mediated by the
scalar field; only gravitational and electromagnetic interactions occur.
The gravitational energy added to the system in equilibrium  - the multi BH solution - by
the massive scalar field is balanced by the electromagnetic energy carried by the field.

%%%%%%%%%%%%%%%%%%%%%%%%%%%%%%%%%%%%%%%%%%%%%%%%%%%%%%%%%%%%%%%%%%%%%%%%%%%
\section{Conclusions}
\label{sec:conclusions}
%%%%%%%%%%%%%%%%%%%%%%%%%%%%%%%%%%%%%%%%%%%%%%%%%%%%%%%%%%%%%%%%%%%%%%%%%%%
A scalar field on a BH geometry does not, generically, admit stationary configurations. In this note
we have showed that an extremal scalar field in the background of a charged, extremal BH geometry
does admit such configurations and we have provided a physical interpretation for them.

Our first observation was that the frequencies of quasi-bound states of a
massive, charged, minimally coupled scalar field in the RN background have a well defined
behaviour when a double extremal limit, for both the test field and the background is
taken: the imaginary part vanishes and the real part becomes equal to the field mass. 
Then we showed that in such double extremal limit, configurations with a real frequency equal to the
particle's mass exist, corresponding to 
a distribution of extremal scalar particles, placed at arbitrary
locations in the exterior of the extremal (multi-)BH
solution. If none of these particles sits at the BH horizon,  the configuration is regular therein.
One may argue, however, that the field is irregular
at the location of the sources. But this is the traditional problem in classical field
theory associated to point-like sources. 

The stationary scalar field states we have exhibited are due to no-force configurations between
scalar sources and extremal BHs, at linear level: the gravitational attraction is being balanced by
the electromagnetic repulsion. At non-linear level, however, the scalar field will back react on the
geometry and, since it is charged, it will source the
Maxwell field wherever the scalar field is non-trivial and not just at the location of the
sources, in contrast to the typical multi-centre solutions found in Supergravity/String theory (see,
e.g. \cite{Youm:1997hw}).  
It would will be interesting, but also challenging, to study the configurations we have analyzed
herein at non-linear level, as solutions of the corresponding Einstein-Maxwell-scalar field theory.

%%%%%%%%%%%%%%%%%%%%%%%%%%%%%%%%%%%%%%%%%%%%%%%
\section*{Acknowledgements}
%%%%%%%%%%%%%%%%%%%%%%%%%%%%%%%%%%%%%%%%%%%%%%%
We would like to thank Jo\~ao Rosa for discussions and comments on this draft. JCD Acknowledges
CONACyT-M\'exico
  support.  This work was also supported by the {\it NRHEP--295189} FP7-PEOPLE-2011-IRSES Grant, and
by
  FCT -- Portugal through the project PTDC/FIS/116625/2010.

\bibliographystyle{h-physrev4}
\bibliography{num-rel}
\end{document}